# Atomic-scale imaging and charge state manipulation of NV centers by scanning tunneling microscopy

Arjun Raghavan*[1,2], Seokjin Bae*[1,2], Nazar Delegan[3,4], F. Joseph Heremans[3,4], and Vidya Madhavan[†1,2]

[1]Department of Physics, University of Illinois Urbana-Champaign, Urbana, Illinois 61801, USA
[2]Materials Research Laboratory, University of Illinois Urbana-Champaign, Urbana, Illinois 61801, USA
[3]Pritzker School of Molecular Engineering, University of Chicago, Chicago, Illinois 60637, USA
[4]Q-NEXT, Argonne National Laboratory, Lemont, Illinois 60439, USA

*Equal contribution
[†]vm1@illinois.edu

## Abstract

Nitrogen-vacancy (NV) centers in diamond are among the most promising solid-state qubit candidates, owing to their exceptionally long spin coherence times, efficient spin-photon coupling, room-temperature operation, and steadily advancing fabrication and integration techniques. Despite significant progress in the field, atomic-scale characterization and control of individual NV centers have remained elusive. In this work, we present a novel approach utilizing a conductive graphene capping layer to enable direct imaging and manipulation of NV⁻ defects via scanning tunneling microscopy (STM). By investigating over 40 individual NV⁻ centers, we identify their spectroscopic signatures and spatial configurations. Our dI/dV conductance spectra reveal the ground state approximately 300 meV below the Fermi level. Additionally, density-of-states mapping uncovers a two-lobed wavefunction aligned along the [111] crystallographic direction. Remarkably, we demonstrate the ability to manipulate the charge state of the NV centers from NV⁻ to NV⁰ through STM tip-induced gating. This work represents a significant advancement in the atomic-scale understanding and engineering of NV centers, paving the way for future quantum device development.

Nitrogen-vacancy (NV) centers in diamond[1–8] are defects consisting of a nitrogen substitution and an adjacent carbon vacancy in the diamond lattice (Fig. 1a). The negatively charged NV centers (NV$^-$) have attracted exponentially growing interest in recent years primarily as candidates for quantum sensing and communication[1,3,9]. However, despite the tremendous interest, a major gap remains in our ability to map out NV$^-$ centers at the atomic-scale and probe the effect of the atomic-scale electrostatic environment around NV$^-$ centers on their qubit properties[10–12]. Critically, manipulation of individual NV charge states at atomic length scales has continued to pose a challenge[11,13,14].

Scanning tunneling microscopy (STM) is one of the ideal probes for atomic-scale imaging, mapping, and manipulation of individual defects and to map their response to their local environment. The study of NV$^-$ centers by STM has thus far not been possible due to the insulating characteristics of the diamond crystal host lattice. Here, we use a combination of characterization techniques to identify and manipulate individual NV$^-$ centers in diamond. Motivated by recent STM measurements on insulators[15–18], we introduce a novel technique in which a monolayer of graphene is transferred onto the diamond surface, creating a conductive interface that enables stable tunneling measurements while preserving the electrostatic environment of the underlying atomic defects. The graphene layer is electronically transparent, allowing us to probe the low energy electronic states of NV$^-$. Finally, all STM experiments were carried out in a newly developed laser-STM system which allows us to use in-situ photoluminescence (PL) as an independent confirmation of the presence of NV$^-$ centers.

The samples used in this study are type-1b diamond crystals with the (100) facet facing the surface normal direction. They were fine polished and strain-released[7] before irradiation with an electron beam at 2 MeV, with a dose of $2\times10^{16}$/cm$^2$ to create the NV$^-$ defects. As shown in the atomic force microscopy (AFM) images from Fig. S1, this leads to very smooth and homogeneous surfaces over large length-scales. We then use a wet transfer technique (see Methods), to cover the diamond sample with monolayer graphene. Raman spectra taken after this procedure (Fig. S2a) show the appearance of graphene G- and 2D-peaks in addition to the 1332 cm$^{-1}$ zone-center phonon peak[19,20] of diamond. The 2D-peak is significantly stronger than the G-peak, confirming the graphene to be monolayer. As seen in the AFM images in Figs. S2b-c, the surface remains flat, with the exception of occasional folds in the graphene layer.

For STM studies, after transferring monolayer graphene we secure the graphene/diamond onto our sample holder using a Mo clamp which serves as the bias electrode. A schematic of our measurement configuration is shown in Fig. 1b. STM images (Fig. 1c) show the honeycomb lattice of the surface graphene layer. A larger area scan of this region is shown in Fig. S3. dI/dV spectroscopy taken over a large energy range displays the expected 5.5 eV band gap from diamond from the valence band edge at -3.2 eV to the conduction band edge at +2.3 eV (Fig. 1d). A smaller energy-range spectrum reveals the expected *V*-shape density of states feature from graphene (Fig. 1d inset). The ability to clearly see the diamond bands demonstrates that the graphene monolayer still allows access to the electronic states from the bulk substrate below.

We first obtain optical evidence of the presence of NV$^-$ defects within our STM scanning field-of-view. To do this, we measure PL spectra at our tip-sample junction using a 532 nm excitation laser with a 5 $\mu$m beam spot diameter. As shown in Fig. 1e, the characteristic PL signal from NV$^-$ defects is found, including the zero-phonon line at ~637 nm (inset shows a camera image of the beam focused at the tip-sample junction). Having confirmed the presence of NV$^-$ centers, we move on to atomic-scale STM characterization of individual defects.

For clarity of presentation, we will hereafter label defects measured by STM sequentially, as Defect 1, Defect 2, etc. A prominent defect that is discerned in negative bias topographies, labeled Defect 1, is shown in Fig. 2a. dI/dV spectroscopy on Defect 1 shows a characteristic peak at approximately -300 meV (below the Fermi energy), as displayed in Fig. 2c. As shown in Fig. S4, the defect peak is robust at various tip-sample junction conditions and a large energy range dI/dV spectrum shows no additional peaks (Fig. S5). We note that the defect cannot be observed in the topography at positive biases (Fig. 2b). This indicates[21–23] that the defect resides beneath the surface—rather than being an adsorbate on the graphene layer—as it does not affect the local topography and is predominantly visible at negative sample biases due to the density of states peak at -300 meV.

To obtain statistical information on these and other defects seen in samples, we measured ~45 defects on two different type-1b diamond samples, with more than 5 different STM tips. The histogram of measured defect peak energies shows two maxima (Fig. 2f). We find that most defects show a peak at ≈-300 meV. Comparing the known NV[-] energy levels from the band edges[24,25] and using the position of the Fermi energy relative to the conduction and valence bands from our spectroscopy data (Fig. 1d), we expect the NV[-] ground state to fall at approximately -260 meV (|g> in Fig. 2d). The representative peak energy (-300 meV) of the majority of the defects is therefore consistent with the expected ground state energy level of an NV[-] center. In addition, a few defects with a peak at ≈+550 meV are also found. These defects are likely single N atoms substituting for C atoms in the diamond lattice (P1 centers). Experimental[26] and theoretical[27] work indicate that P1 centers should show an energy level 1.7 eV below conduction band edge which would be ≈+600 meV above the Fermi energy in our case (Fig. 2e). We note that our identification of these two types of defects as intrinsic to the diamond is supported by additional considerations. Based on prior work[28–30], C vacancies and N adatoms in graphene are expected to show peaks at different energies. Furthermore, measurements of an electronic grade diamond sample with significantly smaller NV[-] concentration determined by PL intensity show an absence of these defects as presented in Fig. S6.

Having identified the most prominent defects in our data as NV[-] centers, we now proceed to map out the density-of-states around one such defect, labeled Defect 4. Fig. 3a shows an NV[-] defect with its corresponding dI/dV spectrum (Fig. 3b). A slice at the Defect 4 peak energy (-300 meV) is shown in Fig. 3c. The electronic states associated with Defect 4 have a double-lobe structure with one lobe stronger than the other. Given the [111] orientation (see Fig. 1a) of NV[-] defects[3,9,31–34] projected to the (100) surface of the diamond crystals along with the two-part structure of NV-centers, the structure seen in our dI/dV maps aligns with the expected dipole orientation of the NV[-] relative to our substrate orientation (see arrows in Fig. 3c and Fig. 3g). These features are consistently observed on multiple defects. Shown in Fig. 3e-h is another NV[-] defect, labeled Defect 5, probed with the same STM tip and on the same diamond sample as Defect 4. Defect 5 shows the same asymmetric two-lobe structure as Defect 4, but is oriented in another projected [111] direction, helping us rule out an STM tip-shape effect as an origin of the two-lobed structure.

Motivated by previous work on STM tip-induced charge-state manipulation in other materials[15,35–37], we next attempt to manipulate the charge state of NV defects via tip-induced gating. Fig. 4a shows Defect 6, an example defect which is to be manipulated. We position our tip above the defect, retract the STM tip 0.5-1.0 nm out of tunneling, and apply a voltage of +5 V to the sample with respect to the grounded tip for 1 min. After going back into tunneling, STM images of the same area no longer show the bright feature associated with Defect 6, as displayed in Fig. 4b. Additionally, dI/dV spectra reveal a transition from the characteristic defect peak to a spectrum with only the background graphene density-of-states, as shown in Fig. 4c. This procedure can be consistently applied to many defects as well as with different types of STM tips, as shown in Fig. 4d-f and Fig. S7. To ensure that the changes in the topography and spectra are not due to tip-changes, for each manipulated defect, we investigate a nearby reference defect within 25-50 nm of the manipulated defect, as shown in Fig. S8. No changes are observed for the reference defects, confirming that our tip manipulation procedure is localized and tip-changes are not responsible for our observations.

A plausible scenario for the physical process occurring during the tip-induced manipulation procedure is a charge-state conversion from $NV^-$ to $NV^0$. When a large positive bias is applied to the sample relative to the tip, the electric field can push an electron out of the near-surface $NV^-$ defect under the tip, converting the charge state from $NV^-$ to $NV^0$. The ground state of the resulting $NV^0$ defect is expected to be more than 2 eV below the Fermi energy[24,25,38,39] where the graphene density-of-states contribution is large compared to the defect density-of-states (as in Fig. S5), which is consistent with the absence of a distinct peak in dI/dV spectra after the tip-induced manipulation process. As shown in Fig. 4, we have found that this manipulation procedure works with both standard W tips as well as plasmon-resonant Au-coated W tips, opening the door to future studies combining STM and near-field optical spectroscopy. Finally, we note that while reversal of the charge state to its original configuration is in principle feasible, it is challenging to demonstrate in this context. The process requires the application of a large negative sample bias, which interacts with the graphene layer (as shown in Fig. S9f), thereby preventing clear observation of the charge state transition.

In summary, we present the first STM imaging and characterization of individual $NV^-$-centers in diamond. We obtain supporting evidence that we are indeed imaging $NV^-$-centers using our in-situ photoluminescence set-up and by comparing samples known to have different $NV^-$ densities. Using STM tip-induced gating, we are able consistently manipulate individual defect charge states, paving the path to atomic-scale engineering of spin-qubits. Since the procedure works with Au-coated tips, which are known to have a plasmon-resonant response[40–44], tip-enhanced, near-field PL studies could potentially provide further confirmation of the local charge states. Given the promise that NV centers hold[1–6,9,24,25,32,45], the findings described in this work mark a critical advance in the atomic-scale understanding and engineering of solid-state qubits. Future possibilities involve integration of optically-detected magnetic resonance capabilities with STM[46–49] to characterize and engineer qubit performance of individual defects at the atomic-scale in technologically optimal electronic-grade diamond host samples.

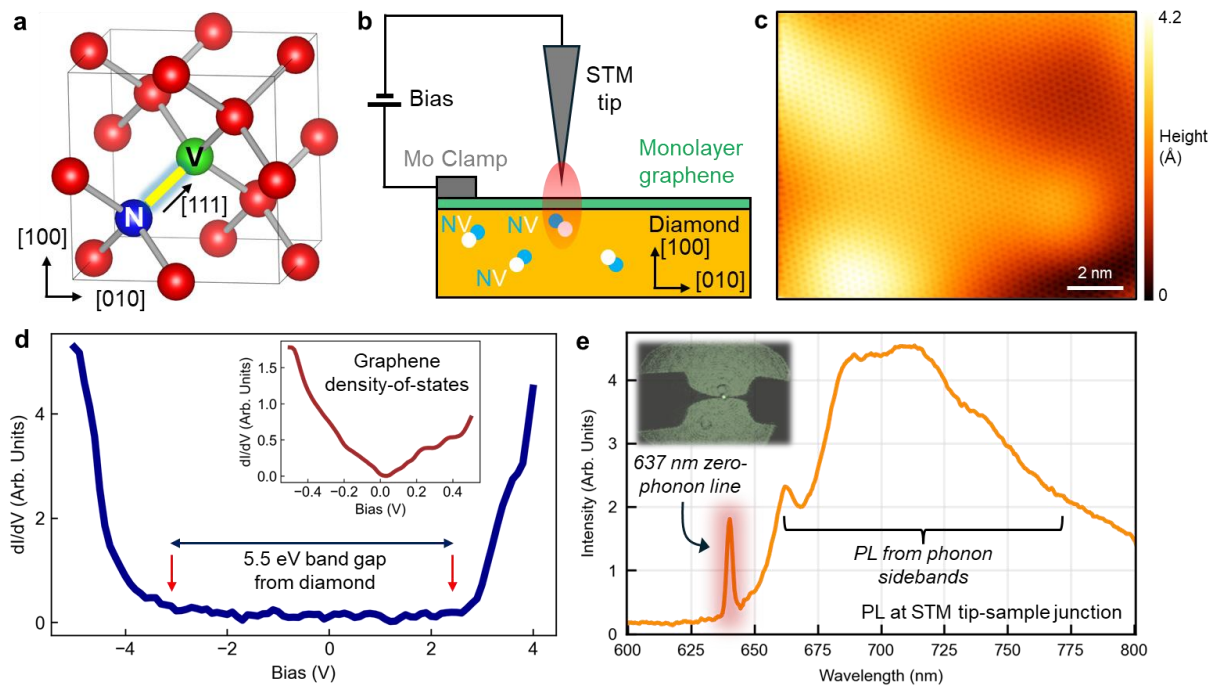

**Fig. 1. Crystal and defect structure, schematic for the device, STM topography and spectroscopy, photoluminescence at tip-sample junction. a,** Crystal structure of diamond, consisting of C atoms (red), along with an NV center, consisting of a N substitution (blue) for C and a neighboring C vacancy (green); NV centers in diamond are oriented along the [111] direction. **b,** Schematic of measurement geometry. Monolayer graphene is transferred onto diamond (100) single crystals and the graphene/diamond structure is clamped down by a Mo plate. Bias voltage is applied through the Mo clamp with tunneling current measurement by an uncoated or Au-coated tungsten STM tip. **c,** STM topography of graphene on diamond showing characteristic honeycomb lattice ($V_{Bias}$ = -500 mV, $I_{Setpoint}$ = 70 pA). **d,** Large energy-range dI/dV spectrum showing 5.5 eV diamond band gap from -3.2 to +2.3 eV ($V_{Bias}$ = -5.00 V, $I_{Setpoint}$ = 70 pA). Inset shows smaller-scale spectrum with expected *V*-shape from graphene ($V_{Bias}$ = -500 mV, $I_{Setpoint}$ = 70 pA). **e,** Photoluminescence from sample in 5 μm diameter laser spot surrounding tip-sample junction (532 nm excitation with 75 μW power, 12s integration, T = 77 K). The expected photoluminescence from NV$^-$ centers, including the characteristic 637 nm zero-phonon line (shaded in red) can be seen clearly, indicating the presence of NV$^-$ centers in our STM field-of-view. The inset is a camera view of the tip-sample junction with the 532 nm laser focused on the tip apex.

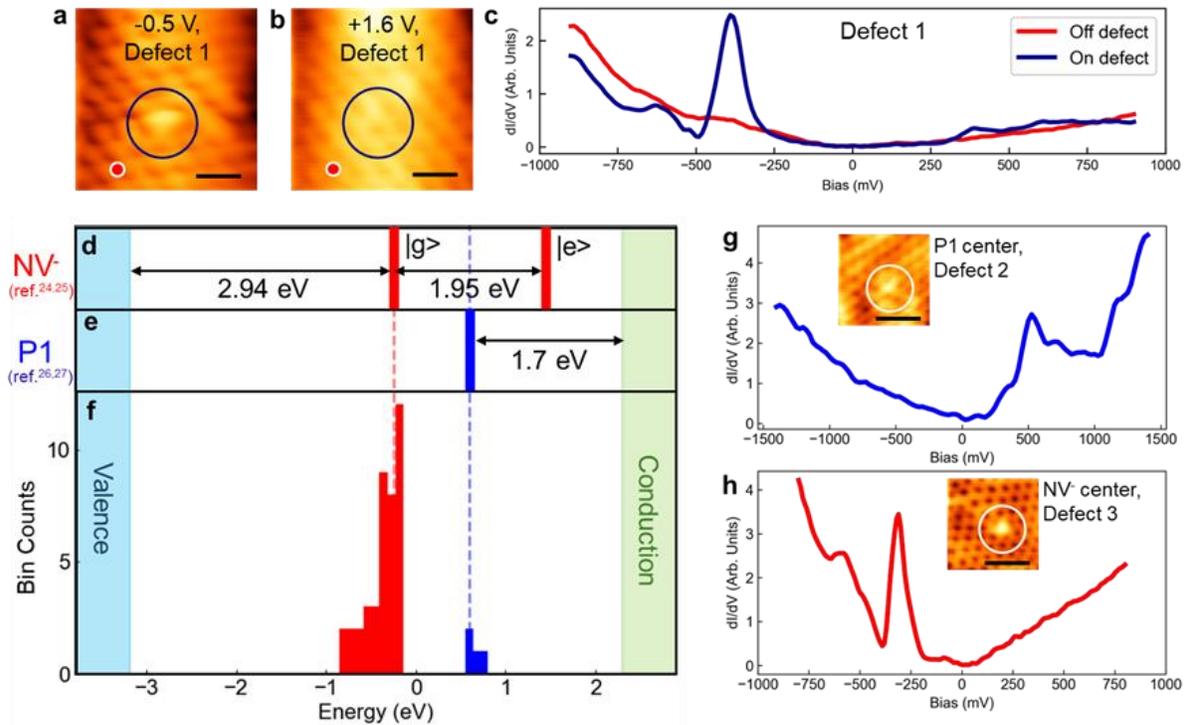

**Fig. 2. STM characterization and energy level distribution of diamond defects. a,** STM topography ($V_{Bias}$ = -500 mV, $I_{Setpoint}$ = 100 pA) of a potential NV$^-$ defect, labeled Defect 1. **b,** Topography from the same area as in **a** at a positive bias ($V_{Bias}$ = +1.60 V, $I_{Setpoint}$ = 70 pA) showing no feature from Defect 1, indicating that the defect is likely located subsurface, in the diamond crystal below the graphene monolayer. **c,** dI/dV spectra ($V_{Bias}$ = -900 mV, $I_{Setpoint}$ = 100 pA) on and away from Defect 1, at the positions marked by the blue circle and red dot shown in **a** and **b**. **d-e,** Energy level diagrams for NV$^-$ and P1 defects in diamond, showing the expected position of the defect states relative to the diamond valence and conduction bands (from Ref[24–27]), as well as the position of the band edges relative to the Fermi energy $E_{Fermi}$ (from Fig. 1d). **f,** Histogram of dI/dV peak energies from measurements of 45+ defects on two different diamond samples by STM. The energies fall into two primary regions. One is centered at approximately -260 meV and the other is centered at +600 meV relative to $E_{Fermi}$. These two peak energies closely match previously reported energy levels of the NV$^-$ ground state[24,25] (marked |g>, with excited state marked |e>) and the P1 center[26,27] (N atom substitution for a C atom) in diamond, respectively. The valence and conduction band positions relative to $E_{Fermi}$ are also shown with sky-blue and green shades. **g,** dI/dV spectrum on a representative P1 center ($V_{Bias}$ = -1.40 V, $I_{Setpoint}$ = 70 pA). Inset shows topography of the P1 defect ($V_{Bias}$ = +1.00 V, $I_{Setpoint}$ = 70 pA). **h,** dI/dV spectrum on a representative NV$^-$ defect ($V_{Bias}$ = -800 mV, $I_{Setpoint}$ = 100 pA). Inset shows a topography of the NV$^-$ defect ($V_{Bias}$ = -1.00 V, $I_{Setpoint}$ = 70 pA). All scale bars in this figure correspond to 0.5 nm.

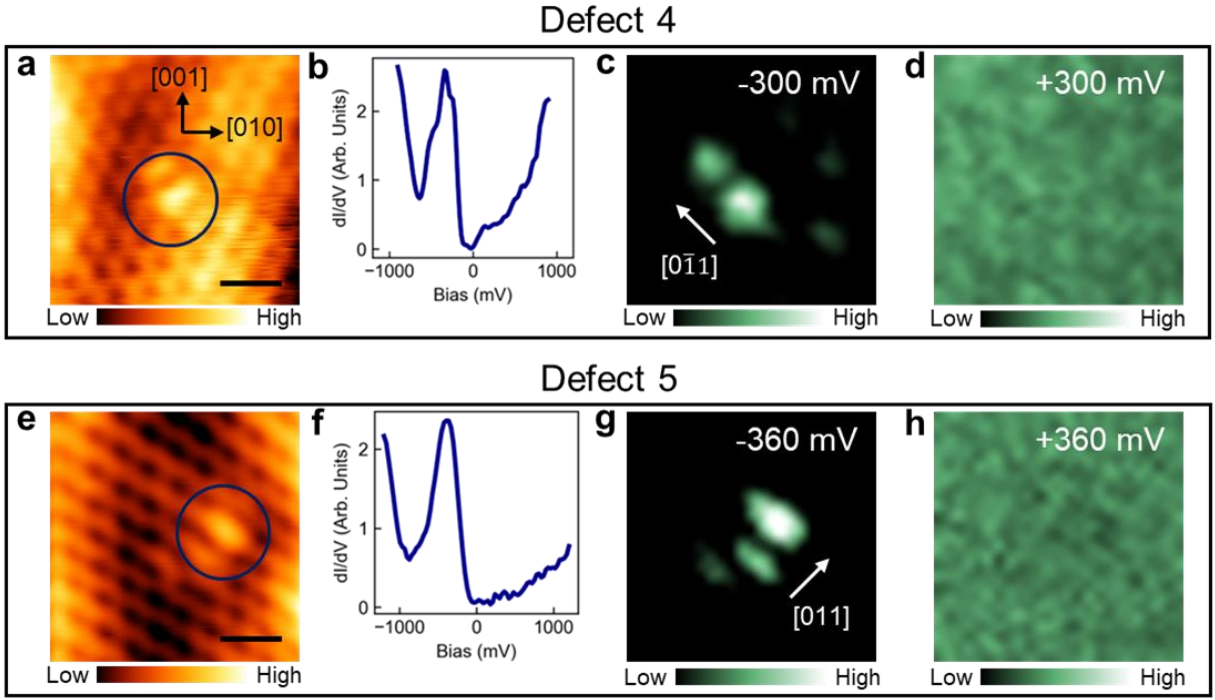

**Fig. 3. Topography, dI/dV spectroscopy, and spatial mapping of NV⁻ defect wavefunctions in diamond. a-d,** Topography ($V_{Bias}$ = -900 mV, $I_{Setpoint}$ = 400 pA), dI/dV spectrum ($V_{Bias}$ = 900 mV, $I_{Setpoint}$ = 100 pA) and two dI/dV map ($V_{Bias}$ = 900 mV, $I_{Setpoint}$ = 100 pA) slices at -300 mV and +300 mV of an NV⁻ defect, labeled Defect 4. A clear asymmetric double-lobe structure can be seen in the density-of-states at the peak energy of -300 mV, while no structure is visible at +300 mV. **e-h** Topography ($V_{Bias}$ = -1.00 V, $I_{Setpoint}$ = 70 pA), dI/dV spectrum ($V_{Bias}$ = -1.20 V, $I_{Setpoint}$ = 70 pA) and two dI/dV map ($V_{Bias}$ = -1.20 V, $I_{Setpoint}$ = 70 pA) slices at -360 mV and +360 mV of another NV⁻ defect, labeled Defect 5, on the same sample and probed with the same STM tip as Defect 4. Again, an asymmetric double-lobe structure can be seen in the negative energy density-of-states at the peak energy of -360 mV but has a different directionality from Defect 4, showing that the spatial structures of the defect wavefunction are not due to tip-shape effects. The dI/dV slice at positive bias again shows no structure. We find that the defects we image are along the diamond [111] direction projected to the (100) surface ([011], [0$\bar{1}$1]), in agreement with the expected directionality of NV-centers. The diamond lattice directions are labeled in panel **a**. All scale bars in this figure are 0.5 nm.

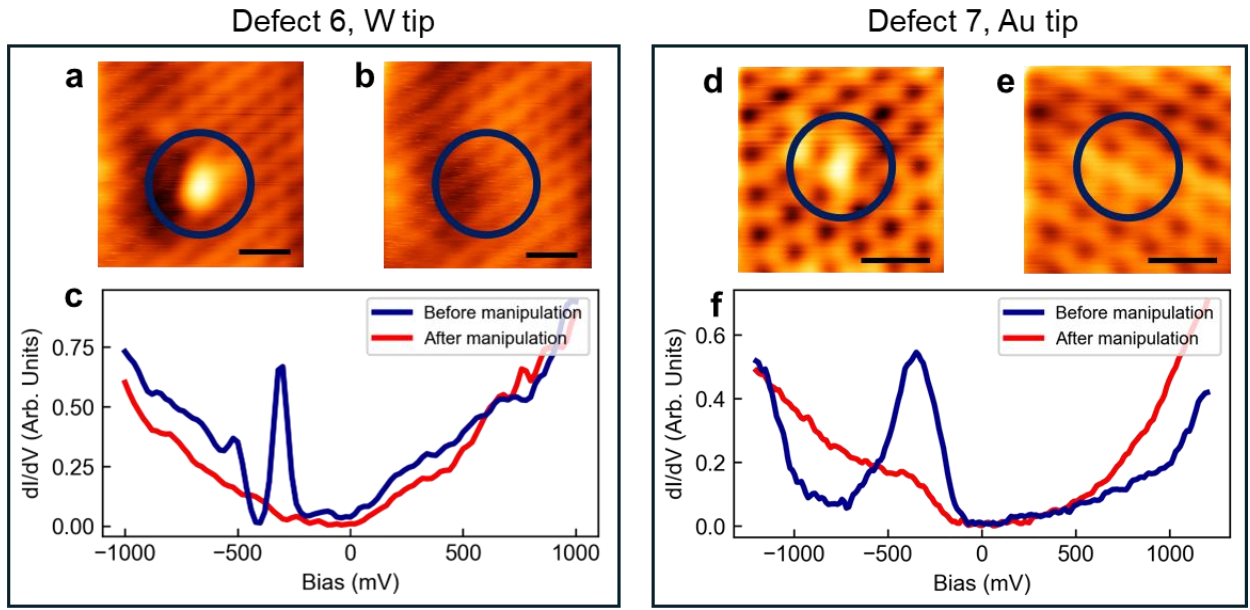

**Fig. 4. STM tip-induced charge-state manipulation of defects in diamond. a-b,** STM topography ($V_{Bias}$ = -1.00 V, $I_{Setpoint}$ = 70 pA) of an NV⁻ defect (labeled Defect 6) (**a**) before and (**b**) after tip-induced gating manipulation of the charge state using a standard W tip by applying +5 V to the sample relative to the tip. **c,** dI/dV spectra ($V_{Bias}$ = 1.00 V, $I_{Setpoint}$ = 70 pA) before and after manipulation of Defect 6 showing the disappearance of the strong defect peak after manipulation. **d-e,** STM topography ($V_{Bias}$ = -900 mV, $I_{Setpoint}$ = 70 pA) of another NV⁻ defect, labeled Defect 7, (**d**) before and (**e**) after charge-state manipulation using an Au-coated W tip, again by applying +5 V to the sample relative to the tip. **f,** dI/dV spectra ($V_{Bias}$ = -1.20 V, $I_{Setpoint}$ = 70 pA) before and after manipulation of Defect 7, again showing the disappearance of the strong defect peak after manipulation. All scale bars in this figure correspond to 0.5 nm.

**Methods**

*STM measurements*: Our STM measurements were performed in a Unisoku USM1200LL instrument at 77 K in ultrahigh vacuum with a Femto DLPCA-200 pre-amplifier. dI/dV measurements are taken with a Stanford Research Systems SR830 lock-in amplifier with a modulation frequency of 911.1 Hz. The sample orientation is determined visually, enabled by optical access to the sample when mounted in the Unisoku USM1200LL STM. STM tips are electrochemically etched in NaOH solutions; uncoated W tips are annealed in vacuum by electron-beam heating before measurements. To prepare Au-coated tips, we use electron-beam evaporation to coat 80 nm of Au onto etched W tips; the coated tips do not have to be annealed in vacuum due to the chemical inertness of Au. FE-SEM images in Fig. S10 show that the tips remain sharp after the Au coating procedure.

*Sample preparation:* Type-1b diamond crystals are purchased from ElementSix and are subsequently strain-released and fine polished. Samples are then irradiated with an electron beam at 2 MeV, with a dose of $2\times10^{16}/cm^2$. Following irradiation, the samples are tri-acid cleaned and annealed at 1200 °C for 2 hours. Electronic-grade crystals are nitrogen-implanted with a moderate dose in a patterned grid yielding ≈16% coverage and a target depth of 5 nm. Monolayer graphene, grown by chemical vapor deposition on both sides of a Cu foil, is purchased from ACS Material, LLC. Polymethyl methylacrylate (PMMA, $C_5H_8O_2$) is spin-coated onto one side of graphene/Cu, following which $O_2$ plasma etching is used to remove graphene from the back side. The Cu is next etched away in ammonium persulfate for 8 hours; the solution is then diluted and the PMMA/graphene film floating in the dilute ammonium persulfate is scooped onto diamond substrates. After baking out the sample at 120 °C for 1 hour, the PMMA is removed by immersion of the sample in acetone for 6 hours. The sample is then annealed for 3 hours at 600 °C in vacuum to remove residues. Following this, the graphene/diamond sample is loaded into the STM prep-chamber through air, after which it is again annealed at 450 °C for 3 hours in ultrahigh vacuum before transferring into the STM for measurements.

*Photoluminescence and Raman measurements:* Room-temperature photoluminescence (PL) and Raman spectra are taken with a Nanophoton Raman11 confocal microscope and 532 nm excitation. A 100× objective lens is used, yielding a diffraction-limited beam spot. PL and Raman spectra from in-situ measurements at the STM tip-sample junction are taken at 77 K, using a StellarNet HyperNOVA spectrometer. Excitation is at 532 nm with a Thorlabs DJ532-40 DPSS laser, and the beam diameter is approximately 5 $\mu m$.

**Acknowledgements:** The authors thank Preetha Sarkar, Haiyue Dong, Michael Altvater, and Karthick Jeganathan for helpful discussions and assistance in sample preparation. This work is primarily funded by Q-NEXT, supported by the U.S. Department of Energy, National Quantum Information Science Research Centers. This work was carried out in part in the Materials Research Laboratory Central Research Facilities, University of Illinois.

**Author Contributions:** A.R. and S.B. conducted STM measurements. S.B. and A.R. constructed the laser-coupled STM setup. N.D. and F.J.H. prepared diamond samples. All authors performed data analysis. A.R., S.B., and V.M. wrote the paper with input from all authors.

**Competing interests:** The authors declare that they have no competing interests.

**Data Availability:** All data presented in this manuscript will be uploaded to the Illinois Databank.

# Supplementary Information

## Atomic-scale imaging and charge state manipulation of NV centers by scanning tunneling microscopy

Arjun Raghavan*[1,2], Seokjin Bae*[1,2], Nazar Delegan[3,4], F. Joseph Heremans[3,4], and Vidya Madhavan[†1,2]

[1]Department of Physics, University of Illinois Urbana-Champaign, Urbana, Illinois 61801, USA
[2]Materials Research Laboratory, University of Illinois Urbana-Champaign, Urbana, Illinois 61801, USA
[3]Pritzker School of Molecular Engineering, University of Chicago, Chicago, Illinois 60637, USA
[4]Center for Molecular Engineering and Materials Science Division, Argonne National Laboratory, Lemont, Illinois 60439, USA

*Equal contribution

[†]vm1@illinois.edu

## Table of Contents

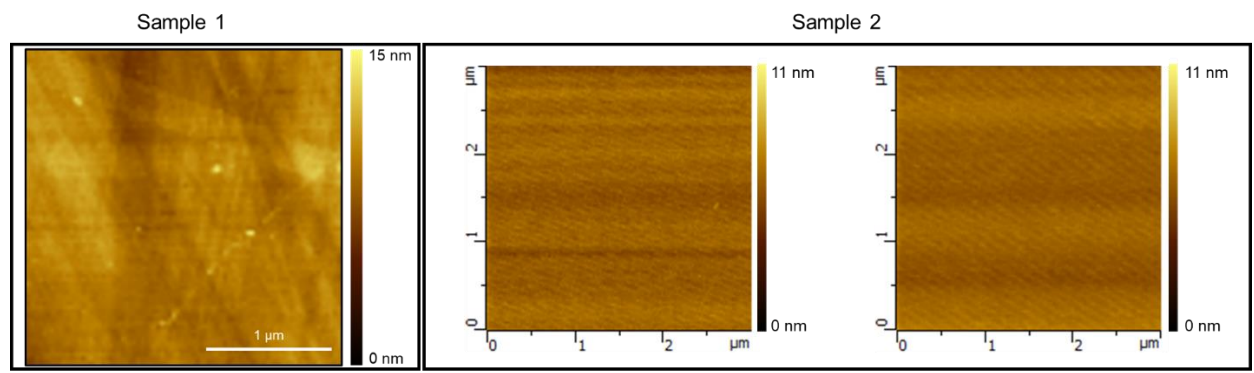

**Fig. S1. AFM images of diamond**. AFM images of two polished and strain-released diamond samples showing flat surfaces before transferring monolayer graphene onto the samples.

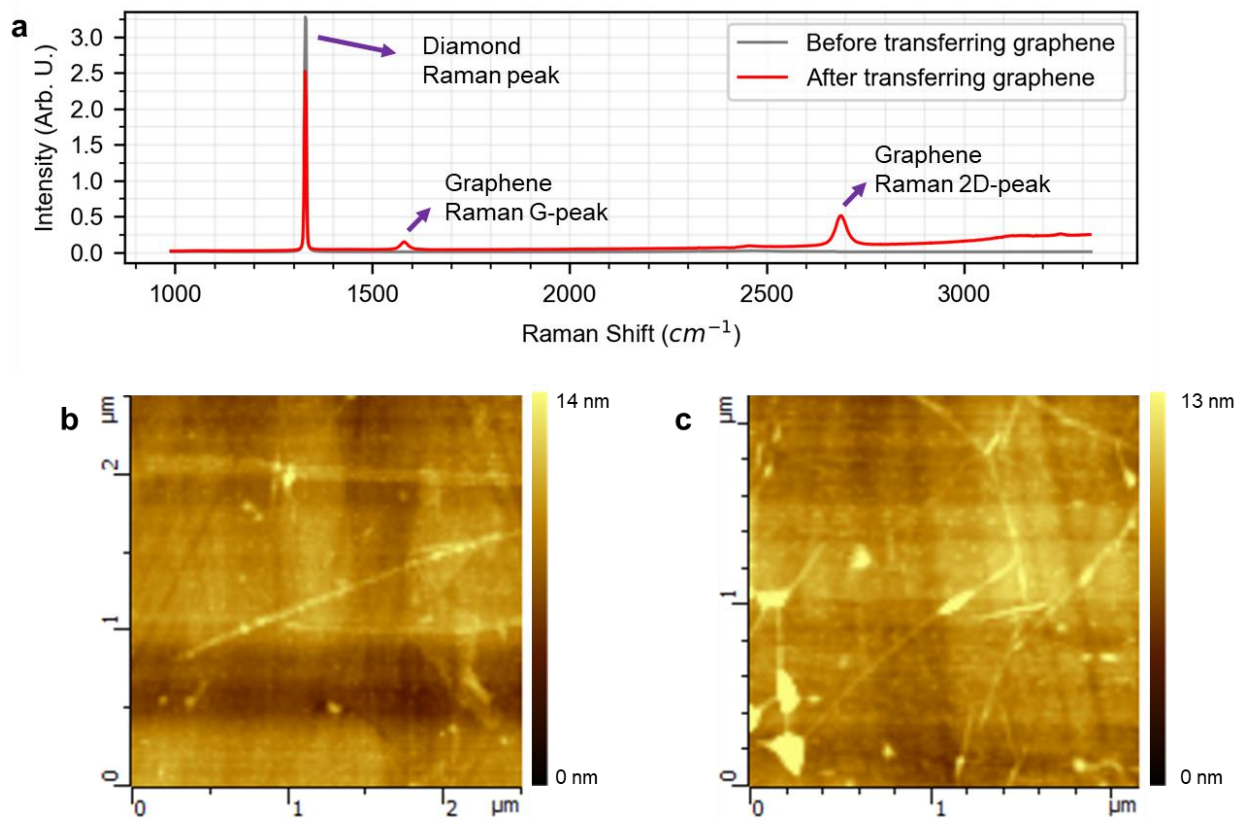

**Fig. S2. Raman spectra comparison and AFM images of graphene on diamond.**
**a,** Raman spectra comparing diamond before and after monolayer graphene is transferred onto it (532 nm excitation with 640 $\mu W$ power, 12.5s integration, T = room temperature). After graphene is transferred, the characteristic Raman G- and 2D-peaks appear, with the 2D-peak significantly stronger than the G-peak, showing the graphene to be monolayer. **b-c,** Representative AFM images from different regions of the diamond sample after graphene is transferred. The sample topography remains flat, with occasional folds from the graphene layer visible.

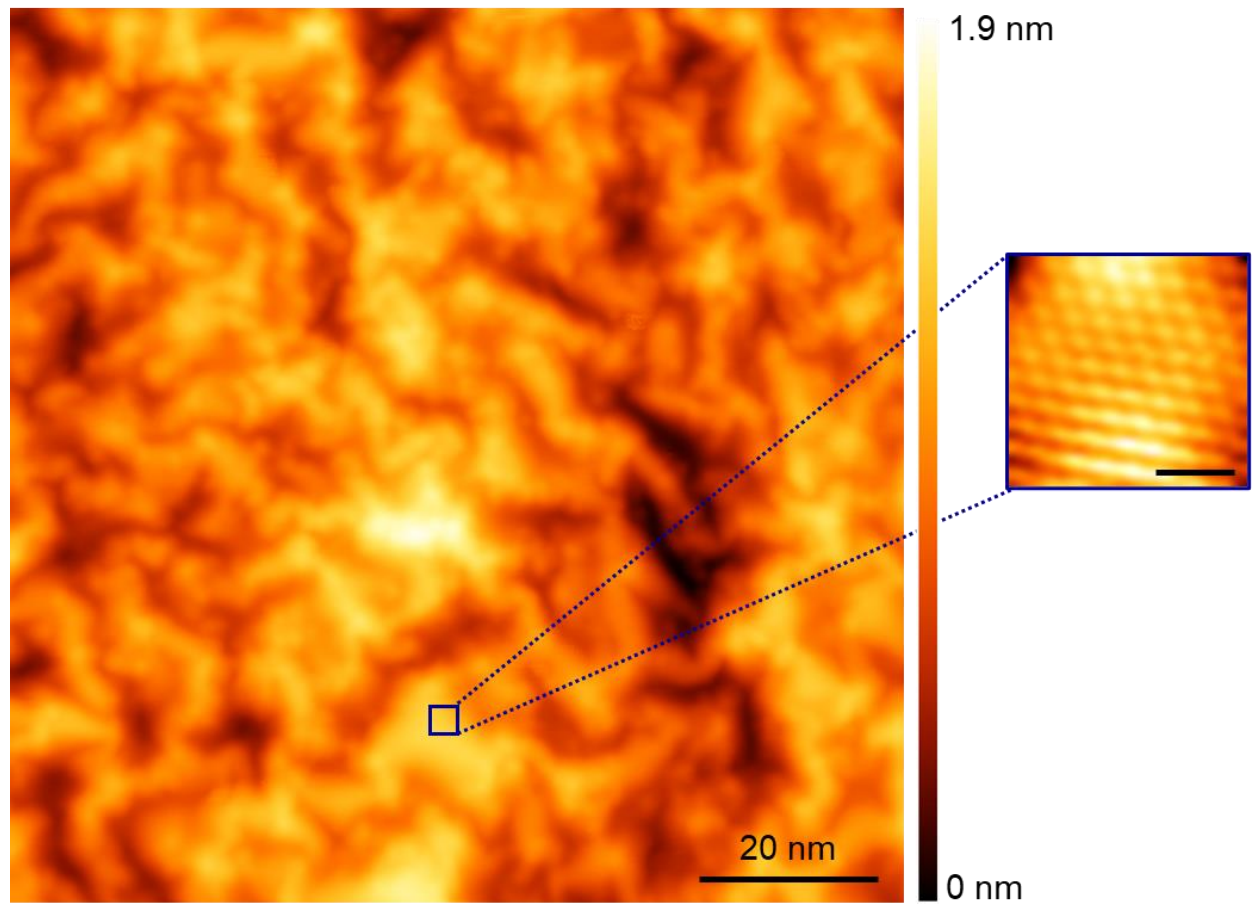

**Fig. S3. Large area topography of monolayer graphene on diamond**. STM topography ($V_{Bias}$ = -1.0 V, $I_{Setpoint}$ = 70 pA) showing large 100×100 nm² area of monolayer graphene on diamond; the total height variation over this area is less than 2 nm. Inset shows atomic resolution topography ($V_{Bias}$ = -1.0 V, $I_{Setpoint}$ = 70 pA) of graphene taken at the same tunneling parameters as the larger scan; scale bar for inset is 0.5 nm.

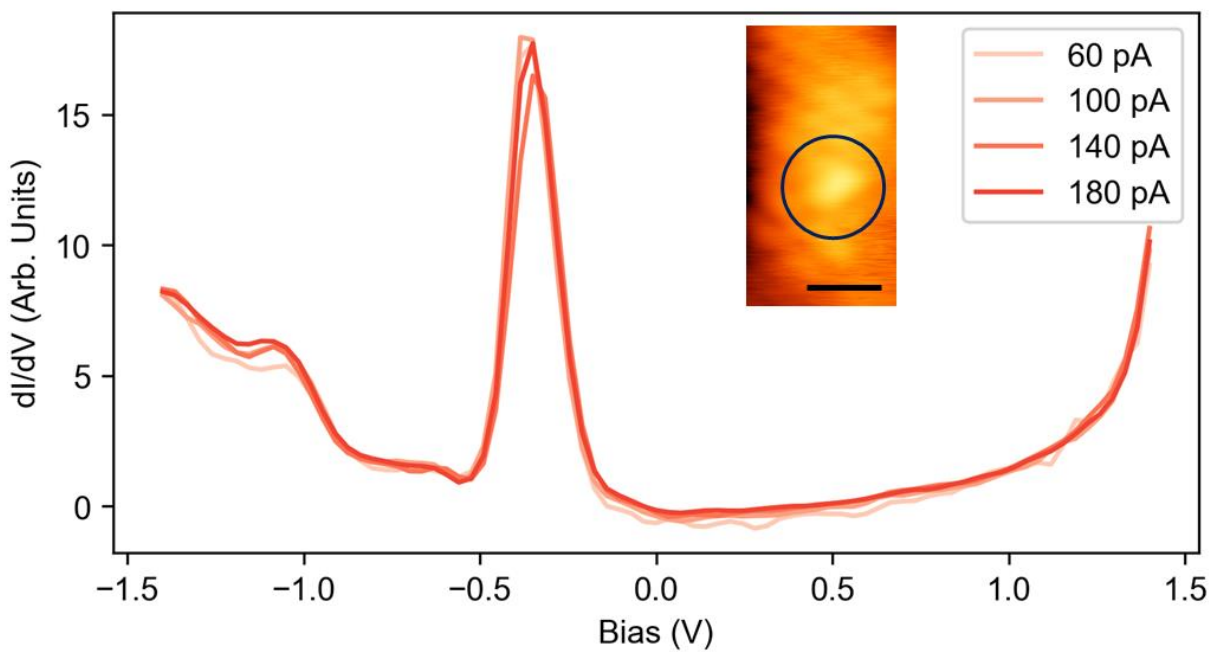

**Fig. S4. Tunneling current setpoint characterization of Defect 1 peak**. dI/dV spectra starting at -1.40 V and finishing at +1.40 V with various tunneling current setpoints on Defect 1. Spectra are normalized by the current setpoint and fully overlap with each other, showing that the defect peak in dI/dV spectra is robust and not dependent on tip-sample distance. Inset shows topography ($V_{Bias}$ = -1.00 V, $I_{Setpoint}$ = 100 pA) of defect; scale bar is 1 nm.

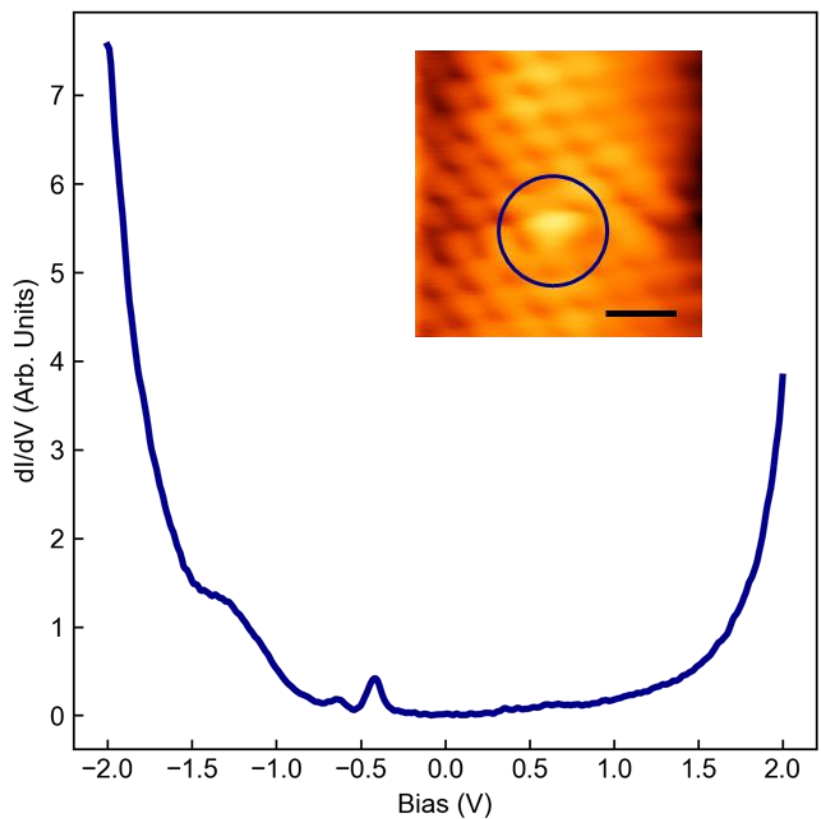

**Fig. S5. Large energy-range dI/dV spectrum on Defect 1.** Large energy-range dI/dV spectrum ($V_{Bias}$ = -2.00 V, $I_{Setpoint}$ = 400 pA) of Defect 1 (topography in inset) shows only a single defect energy level peak over the full -2.00 eV to +2.00 eV energy range. Inset topography is at $V_{Bias}$ = -500 mV and $I_{Setpoint}$ = 100 pA; scale bar is 0.5 nm.

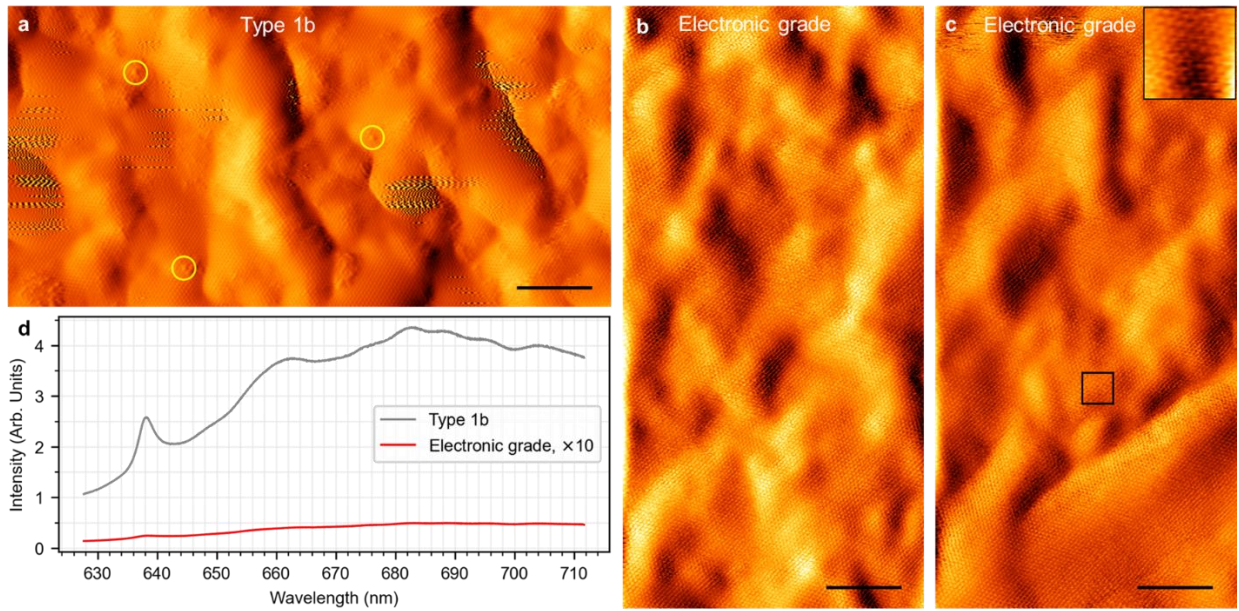

**Fig. S6. Comparison of type-1b and electronic grade diamond samples**. **a,** Derivative image of topography ($V_{Bias}$ = -1 V, $I_{Setpoint}$ = 70 pA) of an area from graphene on a type-1b diamond sample. Yellow circles mark three prominent isolated defects in the image. **b-c,** Derivative images of topographies ($V_{Bias}$ = -500 mV, $I_{Setpoint}$ = 300 pA) of areas from an electronic-grade diamond sample covered with monolayer graphene. No defects can be seen visually; inset in **c** is a zoomed-in 2×2 nm$^2$ topography from the area marked by the black square. **d,** Photoluminescence spectra (532 nm excitation with 520 $\mu W$ power, 5s integration, T = room temperature) from type-1b and electronic grade diamond (multiplied by 10 for visibility); the electronic grade spectrum intensity is ~90 times weaker than the type-1b spectrum intensity. All scale bars in this figure are 5 nm.

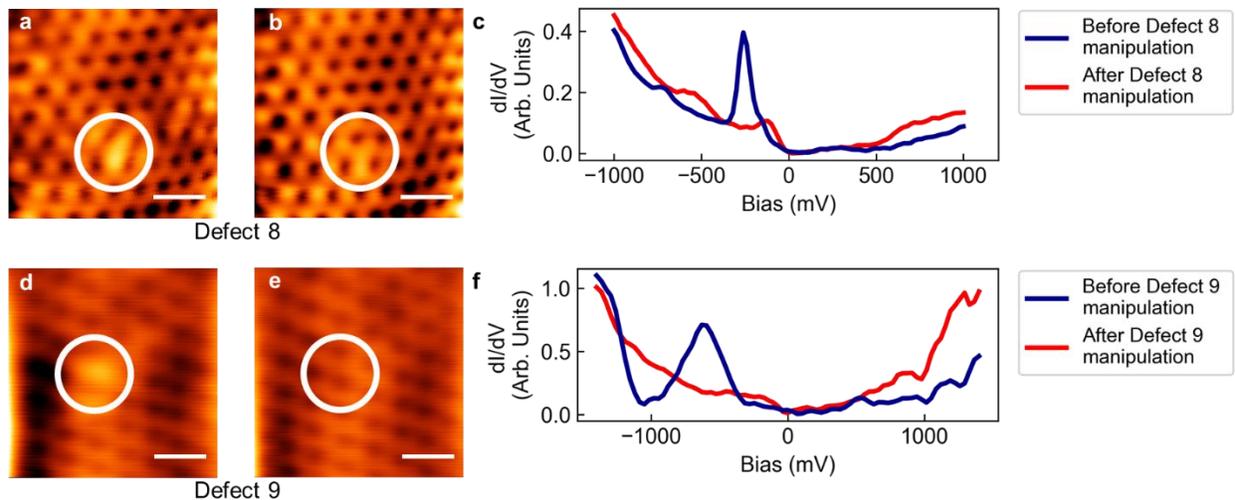

**Fig. S7. Further examples of tip-induced charge-state manipulation of diamond defects with a W tip**. **a,** Topography ($V_{Bias}$ = -400 V, $I_{Setpoint}$ = 70 pA) before tip-induced charge-state manipulation of Defect 8. **b,** Topography ($V_{Bias}$ = -400 mV, $I_{Setpoint}$ = 70 pA) on the same area as **a** after manipulation shows the disappearance of the defect. **c,** Comparison of the dI/dV spectra ($V_{Bias}$ = -1.00 V, $I_{Setpoint}$ = 70 pA) before and after manipulation of Defect 8 shows the disappearance of the defect peak. **d-e,** Topographies ($V_{Bias}$ = -1.50 V, $I_{Setpoint}$ = 70 pA) of another defect, labeled Defect 9, with a different W tip before (**d**) and after (**e**) manipulation of the defect. **f,** Comparison of the dI/dV spectra ($V_{Bias}$ = -1.40 V, $I_{Setpoint}$ = 70 pA) before and after manipulation of Defect 9 shows the disappearance of the defect peak. All scale bars in this figure are 0.5 nm.

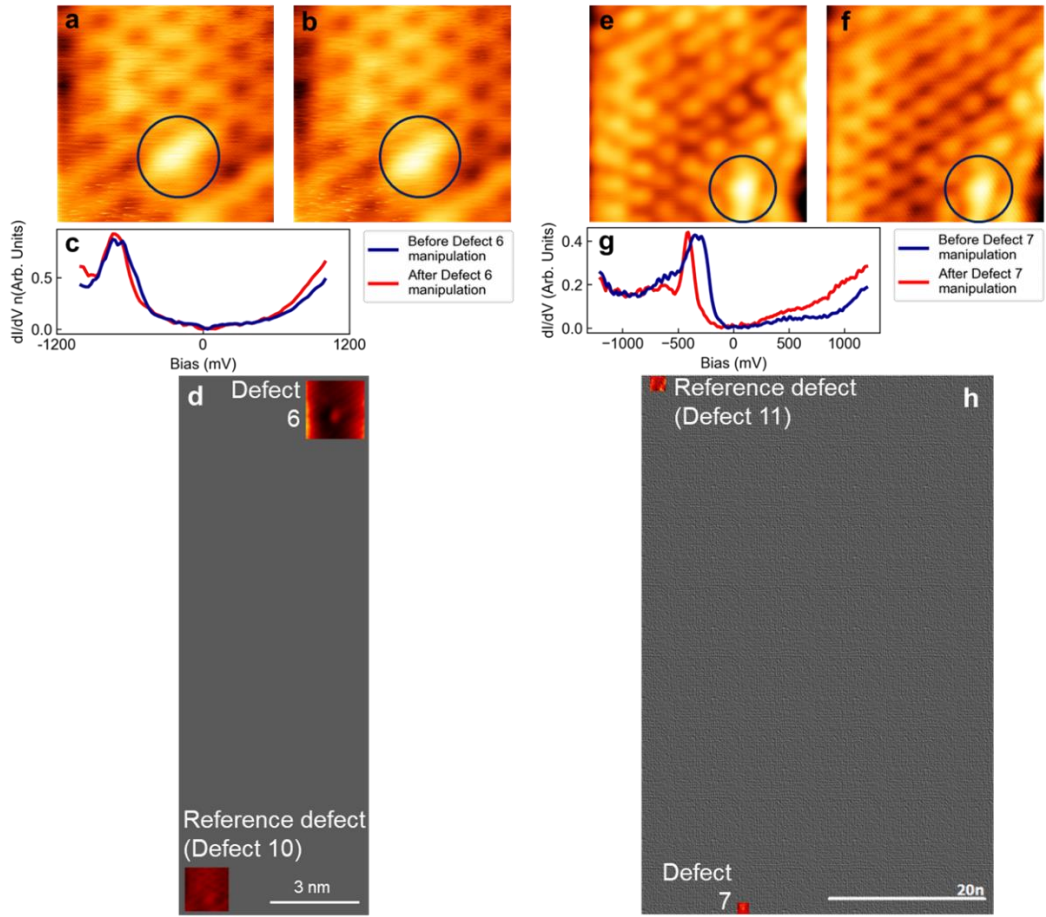

**Fig. S8. dI/dV spectra of the reference defects before/after the tip-induced charge state manipulation and their relative position to the manipulated defects in Fig. 4**. **a,** Topography ($V_{Bias}$ = -1.00 V, $I_{Setpoint}$ = 70 pA) of a reference defect (Defect 10) before manipulation of Defect 6. **b,** Topography ($V_{Bias}$ = -1.00 V, $I_{Setpoint}$ = 70 pA) of Defect 10, after the manipulation of Defect 6. It can be seen that the topographies appear identical for this reference defect. **c,** dI/dV spectra ($V_{Bias}$ = -1.00 V, $I_{Setpoint}$ = 70 pA) Defect 10 before and after manipulation of Defect 6 both show defect peak, indicating that the reference defect has not changed due to the manipulation of Defect 6. **d,** Stitched image showing the relative positions of Defect 6 and its reference defect, Defect 10. The defects are less than 25 nm apart, showing the STM tip-induced manipulation procedure to be highly localized. **e,** Topography ($V_{Bias}$ = -500 mV, $I_{Setpoint}$ = 70 pA) of a reference defect (Defect 11) before manipulation of Defect 7. **f,** Topography ($V_{Bias}$ = -500 mV, $I_{Setpoint}$ = 70 pA) of Defect 11, after the manipulation of Defect 7. It can be seen that the topographies again appear identical for this reference defect. **g,** dI/dV spectra ($V_{Bias}$ = -1.20 V, $I_{Setpoint}$ = 70 pA) on Defect 11 before and after manipulation of Defect 7 both show defect peak, indicating that the reference defect has not changed due to the manipulation of Defect 7. **h,** Stitched image showing the relative positions of Defect 7 and reference defect, Defect 11.

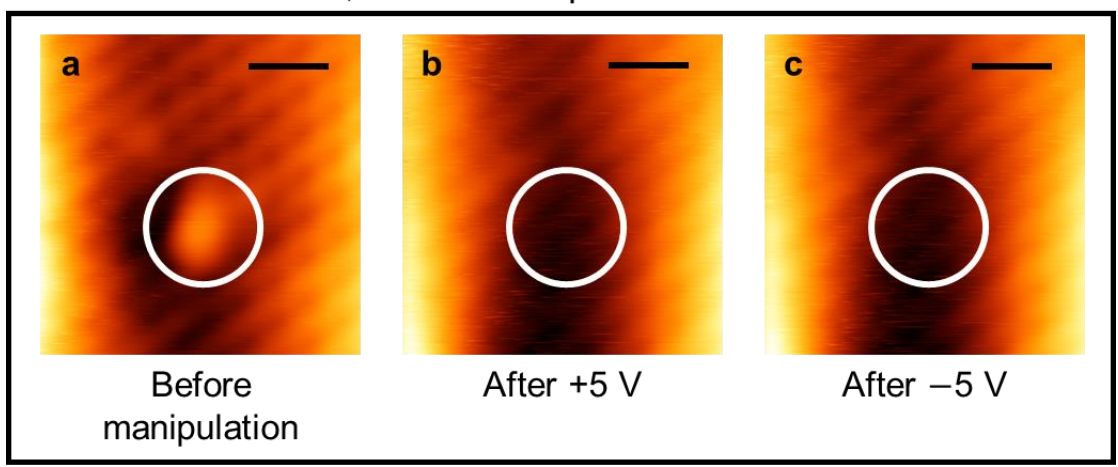

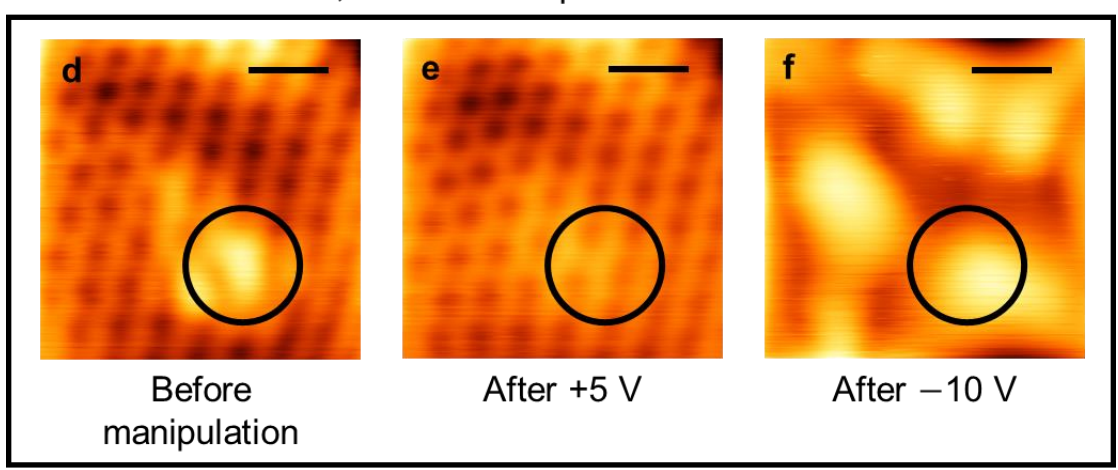

**Fig. S9. Trials to reverse charge state with negative bias. a,** STM topography of a defect, labeled Defect 12, before charge-state manipulation. **b,** Topography after manipulation of charge state of Defect 12 by applying +5 V to the sample relative to the tip shows disappearance of the defect state. **c,** Topography of Defect 12 area after applying -5 V to the sample relative to the tip; no change in the topography is seen. With -5 V, the defect charge-state doesn't return to its original configuration. **d,** Topography of a different defect, labeled Defect 13, before charge-state manipulation. **e,** Topography after manipulation of charge state of Defect 13 by applying +5 V to the sample relative to the tip; the defect state again is shown to disappear in the topography. **f,** Since -5 V didn't have an effect, as shown in panel **c**, the bias was ramped to -10 V. During this procedure however, the surface in the vicinity became highly corrugated and no longer showed an atomically resolved lattice. All scale bars in this figure are 0.5 nm; all topographies are taken at $V_{Bias}$ = -1.00 V and $I_{Setpoint}$ = 70 pA.

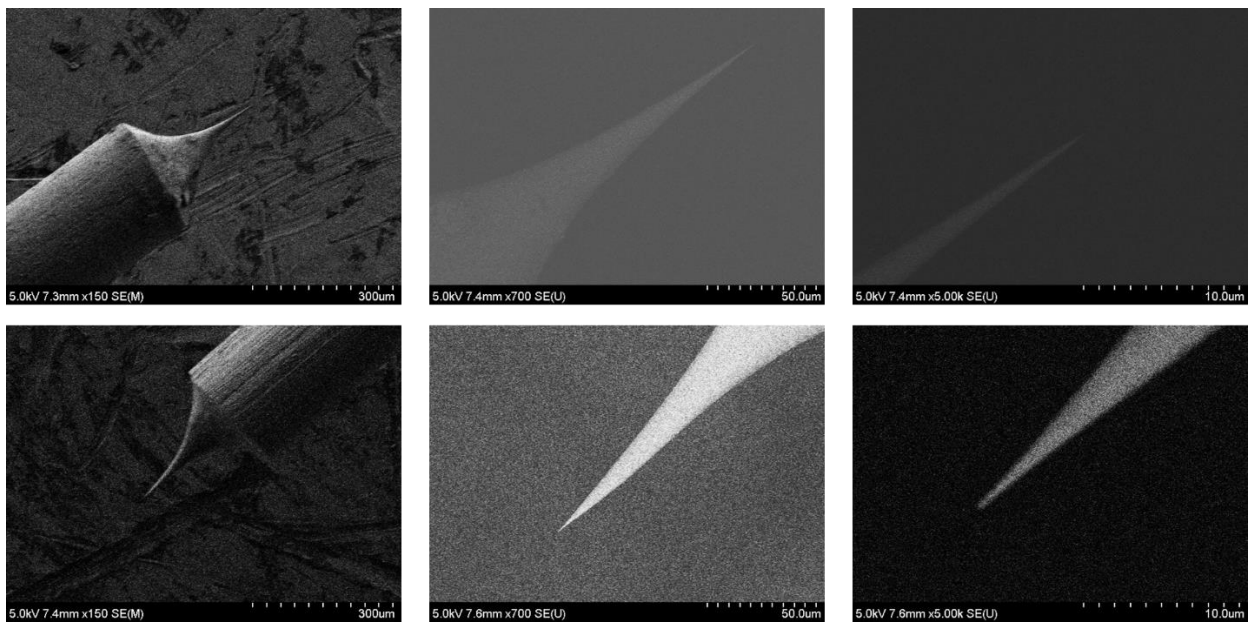

**Fig. S10. FE-SEM images of Au-coated W tips**. Field-emission scanning electron microscopy images (using a Hitachi S-4800 FE-SEM instrument) of two different 80 nm Au-coated W tips at different length-scales show that the tips remain sharp after coating by electron-beam evaporation of Au. The Au-coated W tips were used to probe and manipulate defects in diamond by STM.